\documentclass[aps,twocolumn,pre,showpacs,eqsecnum]{revtex4-1}
\usepackage{graphpap}
\usepackage[dvips]{graphicx}
\usepackage[dvips]{graphics}
\usepackage{color}
\usepackage{amssymb}

\begin{document}

\title{Tunable capacitive inter-dot coupling in a bilayer graphene double quantum dot}

\author{%
  S. Fringes$^1$,
  C. Volk$^{1,2}$,
  B. Terr\'{e}s$^{1,2}$,
  J. Dauber$^{1,2}$,
  S. Engels$^{1,2}$,
  S. Trellenkamp$^{1,2}$, and
  C. Stampfer$^{1,2}$}

% S. Fringes$^1$, B. Terr\'es$^{1,2}$, J. Dauber$^{1}$, C. Volk$^{1,2}$, S. Trellenkamp$^2$, U. Wichmann$^1$, and C. Stampfer$^{1,2}$}
 \affiliation{
$^1$JARA-FIT and II. Institute of Physics B, RWTH Aachen University, 52074 Aachen, Germany, EU\\
$^2$Peter Gr\"unberg Institut (PGI-9), Forschungszentrum J\"ulich, 52425 J\"ulich, Germany, EU
}

\date{ \today}

 \begin{abstract}
We report on a double quantum dot which is formed in a width-modulated etched bilayer graphene nanoribbon. A number of lateral graphene gates enable us to tune the quantum dot energy levels and the tunneling barriers of the device over a wide energy range. Charge stability diagrams and in particular individual triple point pairs allow to study the tunable { capacitive} inter-dot coupling energy as well as the spectrum of the electronic excited states on a number of individual triple points. We extract a mutual capacitive inter-dot coupling in the range of 2~-~6\,meV and an inter-dot tunnel coupling on the order of 1.5\,$\mu$eV.
 \end{abstract}

 %\vspace{0.1cm}
% \pacs{xxxxxxxxxxxxxxxxxxx}
 \maketitle

%%%%%%%%%%%%%%%%%%%%%%%%%%%%%%%%%%%%%%%%%%%%%%%%%%%%%%%
\section{Introduction}
%%%%%%%%%%%%%%%%%%%%%%%%%%%%%%%%%%%%%%%%%%%%%%%%%%%%%%%

Graphene is a promising material for future quantum devices and quantum information technology~\cite{trau07}. In particular, graphene quantum dot devices are interesting candidates for
spin qubits with long coherence times. The weak spin-orbit coupling~\cite{min06,hue06} and weak hyperfine interaction is
expected to be of advantage compared to state-of-the-art GaAs-devices. However, the gapless band structure of graphene makes it hard to electrostatically confine charge carriers. This can be overcome by nanostructuring graphene, where it has been shown that an etching-based "paper cutting" technique leads to graphene nanodevices where transport is dominated by a disorder-induced energy gap. Consequently, graphene nanoribbons~\cite{chen07,han07,mol09,stam09,tod09,liu09,gal10,han10,ter11}, single-electron transistors (SETs)~\cite{sta08,stam08nano,ihn10}, graphene quantum dots (QD)~\cite{gue08,pon08,sch09} even with charge detectors~\cite{gue08,gue11,wan10a} and graphene double quantum dots~\cite{mol09a,mol10a,liu10,wan10,Volk2011} have been demonstrated successfully.
These devices allowed the experimental observation of excited states~\cite{sch08,mol10a}, spin states~\cite{gue10} and the very few carrier regime~\cite{gue09}.
Almost all of these studies were based on single-layer graphene and showed device limitations related to the presence of edge roughness, disorder and unintended vibrational degrees of freedom~\cite{mas10}. Bilayer graphene is an interesting candidate to overcome some of these shortcomings.
In particular, bilayer graphene allows to open a band gap by an out-of-plane electric field~\cite{oos08,zha09}, which may enable a
soft confinement potential and may reduce the influence of localized edge states. Furthermore, it is expected that ripples and substrate-induced disorder are reduced, which may also increase the mechanical stability suppressing unwanted vibrational modes.\\
Here we present a bilayer graphene double quantum dot with a number of lateral gates. The gates allow to tune transport over a wide energy range
and they enable to operate the device in different configurations. We focus on the double quantum dot (DQD) configuration and show characteristic honeycomb-like charge stability diagrams. More specifically, we discuss the electrostatic tunability of the mutual capacitive inter-dot coupling energy. Finally, we also extract { an} inter-dot tunnel coupling energy.

\begin{figure}[t]
\includegraphics*[keepaspectratio=true,width=\linewidth]{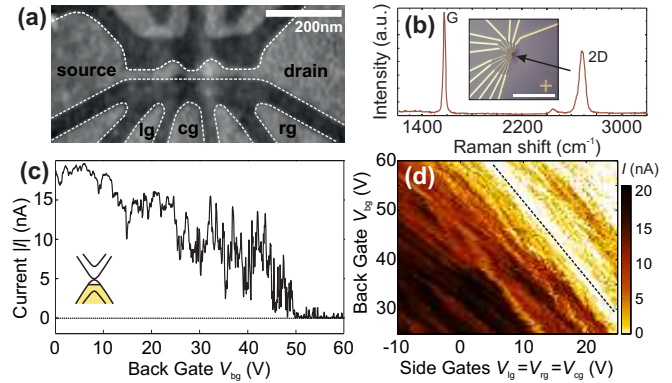}
\caption{%
	(a) Scanning force microscope image of the measured device. The dashed lines highlight the contours of the graphene (light gray).
	(b) Raman spectrum recorded on the used bilayer graphene flake. The inset shows an optical microscope image of the contacted bilayer graphene device (the arrow points to the quantum device and {the} white scale bar is 20\,$\mu$m).
	(c) Back gate characteristics recorded at $V_{\rm{b}} = -20$\,mV.
	(d) Source-drain current as a function of back gate voltage ($V_{\rm{bg}}$) and voltage applied to the side gates ($V_{\rm{lg}}$=$V_{\rm{rg}}$=$V_{\rm{cg}}$) at $V_{\rm{b}}= -20$\,mV (after charge rearrangement).
	}
\label{fig1}
\end{figure}

%%%%%%%%%%%%%%%%%%%%%%%%%%%%%%%%%%%%%%%%%%%%%%%%%%%%%%%
\section{Device Fabrication}
%%%%%%%%%%%%%%%%%%%%%%%%%%%%%%%%%%%%%%%%%%%%%%%%%%%%%%%
The device is fabricated by mechanical exfoliation of natural graphite by adhesive tapes~\cite{nov04,nov05}. The substrate material consists of highly doped silicon (Si$^{++}$) bulk material covered with around 300\,nm of silicon oxide (SiO$_2$) in order to identify bilayer graphene flakes due to absorption and interference. Before deposition, reference alignment markers are patterned on the substrate to relocate the flakes for further processing and characterization. In particular Raman spectroscopy measurements are used to identify bilayer graphene flakes~\cite{fer06,graf07}.
The graphene flakes have to be structured to submicron
dimensions in order to fulfill the device design requirements.
We use a technique based
on resist spin coating, electron beam lithography (EBL),
development and subsequent etching of the unprotected
graphene. We use an EBL resist
[polymethylmethacrylate (PMMA)] with a thickness of
80\,nm and short etching time to define the nanoscale structures.
We follow earlier work~\cite{sta08}, where it has been shown that
short (5 s) mainly physical reactive ion etching (RIE)
based on argon and oxygen (80:20) provides good results
without influencing the overall quality of the graphene flake~\cite{mol07}. After etching and removing the residual EBL resist, the
graphene nanostructures are contacted by an additional
EBL step, followed by metalization and lift-off. Here we
evaporated 5 nm chrome (Cr) and 50 nm gold (Au) for
contacting the bilayer graphene double quantum dot device.\\
A scanning force microscope image of the investigated device is shown in Figure~\ref{fig1}(a), where the graphene structures (bright areas) resting on SiO$_2$ (dark areas) are highlighted by the white dashed lines. The bilayer double quantum dot consists of three 30\,nm wide and 50 to 100\,nm long constrictions, two are separating the source and drain leads from the double quantum dot and one is separating the left (LQD) from the right (RQD) quantum dot. Both bilayer graphene quantum dots have a diameter of approximately 50\,nm. In this study we use the left (lg) and the right (rg) gate to change the number of carriers in the left and the right QD, while the central gate (cg) is used to tune the inter-dot coupling. Additionally, the back gate (bg) is used to adjust the overall Fermi level. The Raman spectrum shown in Figure 1(b) shows that the investigated device is indeed based on a bilayer graphene flake (see also Fig. 1(b) in Ref.~\cite{Volk2011}). The FWHM of the 2D-peak is 57.9\,cm$^{-1}$ with a 18.8\,cm$^{-1}$ spacing of the two inner Lorentzian peaks which in combination with the shoulder on the left side (see 2D peak in Figure 1(b)) provides a good finger print for the bilayer nature~\cite{fer06,graf07}. The inset in Figure~1(b) shows an optical image of the contacted device (see arrow).

%%%%%%%%%%%%%%%%%%%%%%%%%%%%%%%%%%%%%%%%%%%%%%%%%%%%%%%
\section{Measurements}
%%%%%%%%%%%%%%%%%%%%%%%%%%%%%%%%%%%%%%%%%%%%%%%%%%%%%%%
The measurements have been performed in a dilution refrigerator at a base temperature of 10\,mK.
We have measured the two-terminal conductance through the bilayer graphene double quantum dot by applying
a symmetric dc bias voltage while measuring the current through the device with a resolution better than 50\,fA.

\begin{figure}[ht]
\includegraphics*[keepaspectratio=true,width=0.85\linewidth]{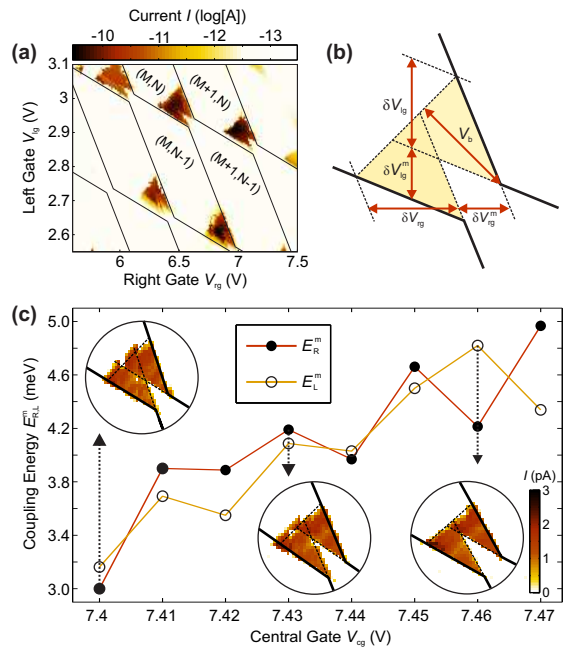}
\caption{
	(a) Source-drain current as a function of the left gate ($V_{\rm{lg}}$) and right gate ($V_{\rm{rg}}$) voltage at an applied bias of ($V_{\rm{b}} = 8$\,mV)
  (b) Illustration of a triple point pair. The quantities are used to deduce the coupling energy.
	(c) Capacitive inter-dot coupling energy as a function of the central gate voltage $V_{\rm{cg}}$ for left (open circle) and right (full circle) gate values. The insets show { three} corresponding triple point pairs.
	}
\label{f:coupling1}	
\end{figure}

\begin{figure*}[t!]%
\centering
\includegraphics*[keepaspectratio=true,width=0.75\textwidth]{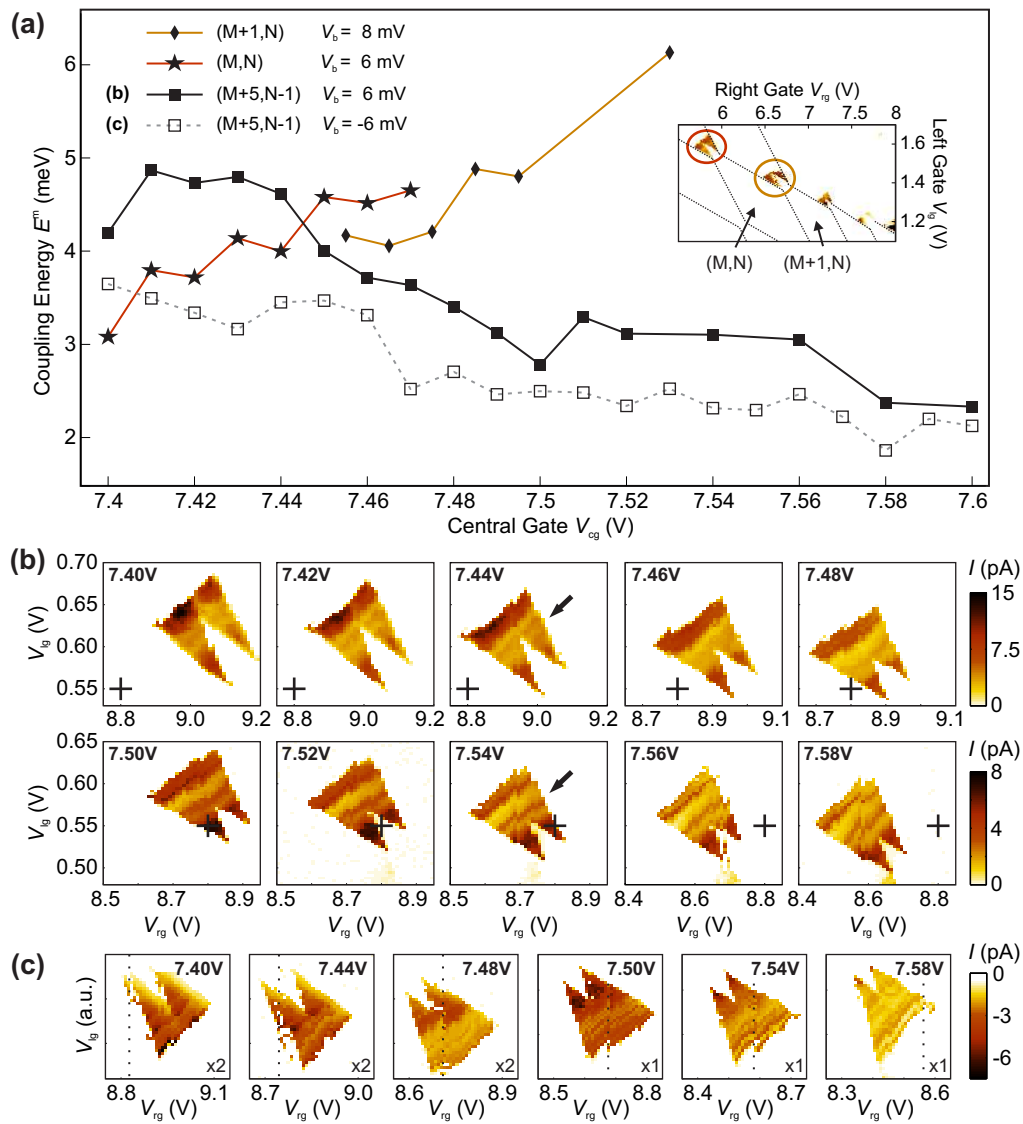}
\caption{
	(a) Mutual capacitive inter-dot coupling energy $E^{\rm{m}}$ plotted as a function of central gate voltage. The corresponding bias and electron configuration (M, N) in the (left, right) quantum dot are given in the upper left corner. The inset shows the position of the triple points in the charge stability diagrams revealing (M, N). The corresponding color scale bar is shown in panel (c) and has to be multiplied by a factor of -0.5.
	(b,c) Current through the double dot as a function of the side gate voltages for different central gate voltages. The very same triple point is measured for $V_{\rm{b}}$=6\,mV (b) and $V_{\rm{b}} = -$6\,mV (c), for	increasing $V_{\rm{cg}}$, ranging from 7.40\,V to 7.58\,V. The capacitive inter-dot coupling energies for (b) and (c) are represented by the black and gray line in panel (a). The black cross in panel (b) highlights the adjustment of the triple point position due to capacitive cross-talk. The dotted line in panel (c) highlights the position of a charge rearrangement. For each row the color scale bar is located on the very right. In the first three triple point pairs in panel (c) the current has to be multiplied by a factor 2.
	}
\label{f:coupling2}	
\end{figure*}

%%%%%%%%%%%%%%%%%%%%%%%%%%%%%%%%%%%%%%%%%%%%%%%%%%%%%%%
\subsection{Device Characterization}
%%%%%%%%%%%%%%%%%%%%%%%%%%%%%%%%%%%%%%%%%%%%%%%%%%%%%%%
To characterize the double quantum dot device on a large energy scale the back gate voltage is tuned over a range of 60\,V. Figure 1(c) shows the source-drain current as a function of the back gate voltage. We observe a so-called transport gap~\cite{stam09} above $V_{\rm{bg}} > 50$\,V ($V_{\rm{b}} = -20$\,mV) and hole dominated transport for lower back gate voltages (see inset in Fig.~1(c)). The large-scale current fluctuations in the transport gap region of the DQD device are due to local resonances in the bilayer graphene constrictions acting as tunneling barriers. In Figure 1(d) we show a measurement of the current as a function of the back gate voltage and the left, right and central gate voltages ($V_{\rm{lg}} = V_{\rm{rg}} = V_{\rm{cg}}$) at constant bias ($V_{\rm{b}} = -20$\,mV).
One characteristic slope is observed, which separates the area of suppressed current (upper right corner) from the area of elevated hole transport (lower left corner). The three tunneling barriers can be tuned almost equally by either the back gate or the sum of the left, right and central gates ($\alpha = \Delta V_{\rm{bg}}\,/\,\Delta V_{\rm{lg,rg,cg}}$ = 1.5), which is in a good agreement with the device design and with earlier studies~\cite{gue09}.\\
By fixing the back gate and the central gate voltage we can record charge stability diagrams by varying the left and right gate voltages independently. In Figure 2(a) we show a finite bias ($V_{\rm{b}}=$ 8\,mV) measurement, where the honeycomb pattern is observed, which is characteristic for the charge stability diagram of a DQD~\cite{Wiel2002}. It is crucial that electrical transport through the DQD is only possible in the case where the energy levels in both quantum dots are aligned within the source-drain bias window. From this measurement we can extract the lever arms and the single quantum dot addition energies, such as %which are shown in Table~1.%
 $E_{\rm{add}}^{\rm{LQD}}=$ 23\,meV and $E_{\rm{add}}^{\rm{RQD}}=$ 13\,meV for the left and right quantum dot, respectively.

%\begin{table}[ht]
%  \caption{Lever arm and addition energies of the charge stability diagram.}
%{\small
%\hfill{}
%\begin{tabular}{|l||c|c||c|}
%\cline{2-4}
%\hline
%& Min & Max & average\\
%\hline
%$\alpha_{\rm{lg}}^{\rm{LQD}}$ & 0.058 & 0.064 & 0.061\\
%$\alpha_{\rm{rg}}^{\rm{RQD}}$ & 0.026 & 0.030 & 0.028\\
%\hline
%$E_{\rm{add}}^{\rm{LQD}}$ & 11.3\,meV & 13.6\,meV & 12.45\,meV\\
%$E_{\rm{add}}^{\rm{RQD}}$ &  9.7\,meV & 15.4\,meV & 12.55\,meV\\
%\hline
%\end{tabular}}
%\hfill{}
%\label{tb:tablename}
%\end{table}

%%%%%%%%%%%%%%%%%%%%%%%%%%%%%%%%%%%%%%%%%%%%%%%%%%%%%%%
\subsection{Capacitive Inter-dot Coupling} % of the two quantum dots}
%%%%%%%%%%%%%%%%%%%%%%%%%%%%%%%%%%%%%%%%%%%%%%%%%%%%%%%
In this section we discuss the effect of the central gate voltage on the capacitive coupling between the left and the right quantum dot (i.e. the capacitive inter-dot coupling). In Figure~\ref{f:coupling1}(b) we show a schematic illustration of a triple point pair highlighting the voltage differences, which allow to determine the mutual capacitive coupling energies, $E^{\rm{m}}_{\rm{rg}}$ and $E^{\rm{m}}_{\rm{lg}}$. These coupling energies can be directly extracted from the lever arms and the illustrated right gate and left gate voltage differences, respectively. For example, we obtain $E^{\rm{m}}_{\rm{rg}}$ from
\begin{equation}
	E^{\rm{m}}_{\rm{rg}} = \alpha_{{\rm{rg,R}}} \cdot \delta V^{\rm{m}}_{\rm{rg}},
\end{equation}
where $\alpha_{\rm{rg,R}} = V_{\rm{b}}/\delta V_{\rm{rg}}$ is the right gate lever arm. A similar expression exists also for left gate voltage differences exclusively.
In Figure~\ref{f:coupling1}(c) the coupling energies $E^{\rm{m}}_{\rm{rg,lg}}$, obtained by the right gate voltage values (full circles) and by the left gate voltage values (open circle), are plotted for eight different central gate voltages for one and the same triple point pair. The three insets show examples of the investigated triple point pair for a weak, medium and stronger inter-dot coupling. It is expected that $E^{\rm{m}}_{\rm{rg}}$ and $E^{\rm{m}}_{\rm{lg}}$ have the same values, but due to limited resolution and partial charge rearrangements the edges of the triangles are not very clear, leading to a scattering of both results.

The average values $E^{\rm{m}} = (E^{\rm{m}}_{\rm{rg}} + E^{\rm{m}}_{\rm{lg}})/2$ as function of the central gate voltage are shown in Figure~\ref{f:coupling2}(a) (see stars $\bigstar$). The mutual capacitive coupling energy is increasing non-linearly as function of the central gate voltage. Next we study the tunability of the capacitive coupling energy for near by
triple point pairs.
Since the absolute number of electrons in the two quantum dots are not known, we focus on relative changes and assume as a starting point M electrons in the right and N electrons in the left dot. For a near by triple point pair with one additional electron in the right dot (M+1,N) also a positive slope for $E^{\rm{m}}$ versus $V_{\rm{cg}}$ is observed (see diamonds $\blacklozenge$ in Fig.~\ref{f:coupling2}(a)). However, for the triple point pair with (M+5,N-1) electrons the capacitive inter-dot coupling energy is decreasing as function of the central gate voltage (see full squares $\blacksquare$ in Fig.~\ref{f:coupling2}(a)).\\
The corresponding triple point pairs from which the coupling energies have been extracted are shown in Fig.~\ref{f:coupling2}(b). With increasing central gate voltage the spacing between the triangle tips is decreasing, which is directly related to the decreasing capacitive coupling energy. A closer inspection of the excited states shows not only a coupling energy dependent reduction of the current within the triangles, but also a change in the coupling of the excited state with an
energy of $\approx$~2\,meV (see line running parallel to the base-line; highlighted by the black arrow). In the first row of Figure~\ref{f:coupling2}(b) the excited state resonance is almost completely suppressed and becomes more clear when increasing the central gate voltage (see second row). Due to capacitive cross-talk the overall position of the triple point pair changes in the left and right gate voltage plane as function of the central gate voltage. This is highlighted by the distance to the fixed black cross (+), which is fixed to the very same position in the gate voltage plane. \\
For a negative bias ($V_{\rm{b}} = -6$\,mV) the same dependence is observed (see open squares $\square$ in Fig.~\ref{f:coupling2}(a) and Fig.~\ref{f:coupling2}(c)). Although the absolute value of the bias is the same, the trace with the open squares has an offset, compared to trace with the full squares. Please note that the corresponding triple point pairs (see Fig.~\ref{f:coupling2}(c))  nicely show the effect of the central gate on a charge rearrangement (see vertical dotted line). In the first two panels the rearrangement is located at the left edge of the triple point. With increasing central gate voltage the triple point pair is moving to lower right and left gate voltage values. For the rearrangement, which is located very close to the right barrier (vertical dotted line), the cross-talk is significantly lower and therefore it moves through the triangles and lies outside for central gate voltages above $V_{\rm{cg}} > 7.58$\,V.\\
Overall, we observe coupling energies ranging from about 2\,meV to 6\,meV. This is by a factor of 2 larger, compared to values from an earlier study on a single-layer double quantum dot~\cite{mol09a}. This difference might be explained by the considerably smaller size of the investigated quantum dots in our work.
More importantly we also observe a rather non-monotonic behavior of the capacitive coupling strength
as function of $V_{\rm{cg}}$ and (M,N). This is most
likely due to a disorder dominated electrostatic landscape in the narrow constriction
connecting the two quantum dots, which is in agreement with recent experiments on graphene
tunneling barriers~\cite{gue11}.

\begin{figure}[tb]
\includegraphics*[width=\linewidth]{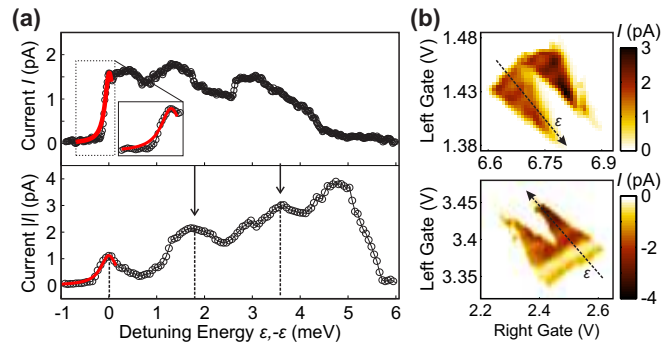}
\caption{
	(a) Line cuts along the detuning energy axis, as shown in panel (b). The bold lines are Lorentzians fitted to the ground state.
	(b) Current as a function of the side gate voltage $V_{\rm{rg}}$ and $V_{\rm{lg}}$ for an applied bias of $V_{\rm{b}}$ = 6\,mV and $V_{\rm{b}}$ = -6\,mV, respectively.
	}
\label{f:tunnelcoupling}	
\end{figure}

%%%%%%%%%%%%%%%%%%%%%%%%%%%%%%%%%%%%%%%%%%%%%%%%%%%%%%%
\subsection{Inter-dot Tunnel Coupling}
%%%%%%%%%%%%%%%%%%%%%%%%%%%%%%%%%%%%%%%%%%%%%%%%%%%%%%%

Transport through the quantum dot states can be analyzed
more quantitatively using the result from Stoof and Nazarov
for resonant tunneling~\cite{sto96}. In the limit of weak inter-dot
tunnel coupling $t_{\rm{m}}\ll\Gamma_{\rm{in,out}}$, where $t_{\rm{m}}$ is the inter-dot tunnel
coupling, and $\Gamma_{\rm{in,out}}$ are incoming and outgoing tunnel rates,
the current $I$ follows a Lorentzian line shape as a function of
the detuning energy $\varepsilon$,
\begin{equation}
I(\varepsilon) = \frac{4et_{\rm{m}}^2/\Gamma_{\rm{out}}}{1+\left(2\varepsilon / h \Gamma_{\rm{out}}\right)^2},
\end{equation}
where $h$ is the Plank constant. In Figure 4(a) we plot two line-cuts along the detuning axis for
a positive and a negative bias ($V_{\rm{b}} = \pm$ 6\,mV), and Lorentzian line fits to
the ground state lines.
Please note that in particular in the lower panel of Fig.~4(a) the excited states at an energy of 1.8\,meV and
3.6\,meV are nicely visible (see arrows).
The corresponding triple points with constant central gate voltages of $V_{\rm{cg}} = 7.465$\,V (upper panel) and $V_{\rm{cg}} = 2$\,V (lower panel) are shown in Figure~\ref{f:tunnelcoupling}(b).
Following Ref.~\cite{liu10} the fitting is done only for the data points
outside of the bias triangle in order to minimize the contributions
from the inelastic transport.
We extract from the
fittings a tunnel rate from the right dot to drain $h\Gamma_{\rm{R} \rightarrow \rm{d}}\approx210\,\mu\rm{eV}$ and an inter-dot tunnel rate $ht_m \approx 1.5\,\mu\rm{eV}$. From negative bias we obtain a tunnel rate from the left dot to source $h\Gamma_{\rm{L} \rightarrow \rm{s}}\approx420\,\mu\rm{eV}$ and an inter-dot tunnel rate $ht_m\approx1.7\,\mu\rm{eV}$.\\
These values are in reasonable agreement with the results reported by Liu et al.~\cite{liu10}. The out of the dot tunnel rates have the same order of magnitude. The inter-dot tunnel rate, however, is in our case by a factor of six smaller ($t_{\rm{m,Liu}}=10\,\mu\rm{eV}$), which might be explained by the different device design. In their case the double dot is formed by two top gates on a 20\,nm wide graphene nanoribbon. The tunneling barriers, induced and controlled by the 40\,nm wide central gate, seem to be more transparent, compared to our side gated 30\,nm wide constriction. Please note that the extracted inter-dot tunnel rate is also consistent with the upper bound of $t\leq20\,\mu\rm{eV}$ estimated by Molitor et al.~\cite{mol10a} for a single-layer graphene double quantum dot.

\section{Conclusion}
We have fabricated a bilayer graphene double quantum dot device, which can be tuned by a number of all-graphene lateral gates. The device has been characterized on a large energy range and its functionality is demonstrated by charge stability diagrams and well-resolved triple points.\\
We have studied the mutual capacitive inter-dot coupling energy $E^m$ and estimated the tunnel coupling energy $t_m$. By varying the central gate voltage we are able to tune $E^m$ systematically from 2 to 6\,meV which is in good agreement with former results for single-layer graphene double quantum dots~\cite{mol10a,liu10}, taking into account the smaller size of the presented quantum dots. Furthermore, the detailed investigation of {individual} triple point pairs has allowed to determine electronic excited state energies and an inter-dot tunnel rate of around $1.5\,\mu\rm{eV}$ which is roughly one order of magnitude lower, compared to recent results reported for single-layer graphene double dots~\cite{mol10a,liu10}. Most likely this difference might be due to the different device designs and architectures. However, further investigations and more statistics are needed to elucidate this issue.\\
 In summary these results give insights into tunable bilayer graphene double quantum dot devices and the ability to tune the capacitive inter-dot coupling energy which is promising for the implementation of spin qubits in future graphene devices.

%In summary, we demonstrated the functionality of a bilayer graphene
%quantum dot with an integrated charge detector based on a nearby nanoribbon.
%We confirm the detection of charging events in
%regimes where Coulomb blockade resonances can hardly be
%measured or resolved because the current levels are below the 10~fA level. In contrast to state-of-the-art
%quantum-point contact charge detectors, we do not make use of
%slopes of quantized conductance plateaus. We rather use local
%resonances in the bilayer graphene nanoribbon to detect charging, very
%similar as it has been used in single-layer graphene devices..
%This technique is considered to play an important role
%for the investigation of future bilayer graphene quantum dots and
%in particular double quantum dots, where spin-quibt states may become
%accessible.

%\begin{acknowledgement}
We thank A. Steffen, R. Lehmann
and U. Wichmann for the help on sample fabrication
and electronics. Discussions with  J. G\"uttinger and S. Dr\"oscher and
support by the JARA Seed Fund and the DFG (SPP-1459 and FOR-912) are gratefully acknowledged.%\end{acknowledgement}

% Use the following code if you wish to generate your bibliography with BibTeX;
% replace the string "pss-demo" below with the name(s) of
% the BibTeX data base(s) you want to use.
% The resulting bibliography-output (the content of the .bbl file)
% must be pasted back into this file before submission.
% Please also include your BibTeX data base file(s) in your submission
% so that we can re-run BibTeX if necessary.
%
%\bibliographystyle{pss}
%\bibliography{pss-demo}
%
% Replace the following example bibliography with your references
% before submission:

\end{document}